\documentclass[aps,prx,twocolumn, groupedaddress]{revtex4-1}
\usepackage[pdftex]{graphicx}
\usepackage{subfigure}
\usepackage{sidecap}

\usepackage{amsmath}
\usepackage{hyperref}
\usepackage{color}

\graphicspath{{pic/}{plot/}}

\usepackage{amsfonts}
\usepackage{amsmath}
\usepackage{bm}                 % bold math

\newcommand{\sslabel}[1]{\label{sec:#1}}
\newcommand{\elabel}[1]{\label{eqn:#1}}
\newcommand{\flabel}[1]{\label{fig:#1}}
\newcommand{\tlabel}[1]{\label{tab:#1}}

\newcommand{\cref}[1]{\ref{cpt:#1}}
\newcommand{\sref}[1]{\ref{sec:#1}}
\newcommand{\eref}[1]{\ref{eqn:#1}}
\newcommand{\fref}[1]{\ref{fig:#1}}
\newcommand{\tref}[1]{\ref{tab:#1}}

\newcommand{\listBegin}{\begin{tabular}{cp{4.5in}}}
\newcommand{\listEnd}{\end{tabular}}

\newcommand{\matrixBegin}[1]{\left[\!\!\left[ \begin{array}{#1}}
\newcommand{\matrixEnd}{\end{array} \right]\!\!\right]}

\newcommand{\beq}[1]{\begin{equation}\elabel{#1}}
\newcommand{\eeq}{\end{equation}}

\newcommand{\beqa}[1]{\begin{eqnarray}\elabel{#1}}
\newcommand{\eeqa}{\end{eqnarray}}

\newcommand{\abs}[1]{\left| #1 \right|}

\def\({\left(}
\def\){\right)}
\def\[{\left[}
\def\]{\right]}

\newcommand{\fsin}[1]{\sin\!\left( {#1} \right)}

\newcommand{\imag}[1]{Im\!\left[ {#1} \right]}
\newcommand{\real}[1]{Re\!\left[ {#1} \right]}

\newcommand{\partderiv}[2]{\frac{\partial{#1}}{\partial{#2}}}
\newcommand{\partderivT}[2]{\partial{#1}/\partial{#2}}

\newcommand{\half}{\tfrac{1}{2}}

\def\kB{k_B}

\usepackage{siunitx}

\DeclareSIUnit\rtHz{$\sqrt{\si{\Hz}}$}
\DeclareSIUnit\mrtHz{\meter\per\rtHz}
\DeclareSIUnit\msolar{{$M_\odot$}}

\definecolor{spring}{rgb}{0.7,0.9,0.7}
\definecolor{brick}{rgb}{0.7,0.2,0.1}
\definecolor{redHL}{rgb}{1.0,0.5,0.5}

\def\mtant{{\mbox{Ta}_2\mbox{O}_5}}
\def\tant{$\mtant$}
\def\sil{$\mbox{SiO}_2$}

\def\kB{k_B}

\def\rG{r_G}

\def\prat{\sigma}

\def\abar{\bar{\alpha}}

\def\bbar{\bar{\beta}}
\def\rbar{\bar{r}}
\def\rhobar{\bar{\rho}}

\def\dcdp{\partderiv{\phi_c}{\phi_j}}
\def\tdcdp{\partial \phi_c / \partial \phi_j}

\def\coat{{c}}

\def\phic{\phi_\coat}

\begin{document}

% Use the \preprint command to place your local institutional report
% number in the upper righthand corner of the title page in preprint mode.
% Multiple \preprint commands are allowed.
% Use the 'preprintnumbers' class option to override journal defaults
% to display numbers if necessary
%\preprint{}

%Title of paper
\title{Multi-Material Coatings with Reduced Thermal Noise}

% repeat the \author .. \affiliation  etc. as needed
% \email, \thanks, \homepage, \altaffiliation all apply to the current
% author. Explanatory text should go in the []'s, actual e-mail
% address or url should go in the {}'s for \email and \homepage.
% Please use the appropriate macro foreach each type of information

% \affiliation command applies to all authors since the last
% \affiliation command. The \affiliation command should follow the
% other information
% \affiliation can be followed by \email, \homepage, \thanks as well.
\author{William Yam}
\author{Slawek Gras}
\author{Matthew Evans}
%\email[]{Your e-mail address}
%\homepage[]{Your web page}
%\thanks{}
%\altaffiliation{}
\affiliation{Massachusetts Institute of Technology, 185
  Albany St, Cambridge, MA 02139, USA}

%Collaboration name if desired (requires use of superscriptaddress
%option in \documentclass). \noaffiliation is required (may also be
%used with the \author command).
%\collaboration can be followed by \email, \homepage, \thanks as well.
%\collaboration{}
%\noaffiliation

\date{\today}

\begin{abstract}
The most sensitive measurements of time and space are made
 with resonant optical cavities, and these measurements are limited by coating thermal noise.
The mechanical and optical performance requirements placed on coating materials,
 especially for interferometric gravitational wave detectors, have proven
 extremely difficult to meet despite a lengthy search.
In this paper we propose a new approach to high performance coatings;
 the use of multiple materials at different depths in the coating.
To support this we generalize previous work on thermal noise in
 two-material coatings to an arbitrary multi-material stack,
 and develop a means of estimating absorption in these multi-material coatings.
This new approach will allow for a broadening of the search for
 high performance coating materials.
 
\end{abstract}

% insert suggested PACS numbers in braces on next line
% insert suggested keywords - APS authors don't need to do this
%\keywords{}

%\maketitle must follow title, authors, abstract, \pacs, and \keywords
\maketitle

% body of paper here - Use proper section commands
% References should be done using the \cite, \ref, and \label commands

%%%%%%%%%%%%%%%%%%%%%%%%%%%%%%%%%%%%
\section{Introduction}
\sslabel{intro}

Lasers and resonant optical cavities have become a ubiquitous tool 
 in optical experiments which explore the bounds of physics through
 precise measurements of space and time
 \cite{McClelland2011, Schnabel2010, Poot2012, Zavattini2012}.
The precision with which these measurements can be made is
 in large part determined by the fundamental thermal motion of the coatings
 used in optical resonators \cite{Kessler2012, Cole2013, Adhikari2014}.

Interferometric gravitational wave detectors in particular set extremely
 stringent requirements on their coatings;
 they must simultaneously have good mechanical properties
 for low thermal noise, low optical absorption for high-power operation,
 and good surface figure to support multi-kilometer resonant cavities
 \cite{Fritschel2014b}.
These requirements are, however, very hard to meet in a single material
 \cite{Crooks2002, Harry2002, Yamamoto02-1, Penn2003, Black2004,
 Crooks2006, Harry2006, Harry07,
 Abernathy2008, Martin2008, Martin2009,Flaminio2010}.

This paper provides the theoretical foundation for a new approach to
 the search for high quality coating materials.
We compute the coating thermal noises and absorption which will result from
 coatings comprised of a variety of materials.
This is to be contrasted with previous works,
 which have assumed that optical coatings are made of one low-index material
 and one high-index material \cite{Hong2013, Evans2008}.

In the next section we present our model of coating Brownian
 and thermo-optic noises, generalized from previous works to allow for
 multi-material coatings.
We go on to develop a simple model of absorption as a function of depth
 in the coating, from which we are able to assess the impact of using
 relatively high absorption materials deep in the coating.
These calculations are followed by a few example of how this new
 approach can be used to produce coatings with lower thermal noise.
All frequently used symbols are given in table \tref{SymbolDefs},
 and appendix \sref{Relate} connects the notation used in this work to that of previous authors.

%%%%%%%%%%%%%%%%%%%%%%%%%%%%%%%%%%%%
\section{Model of Coating Thermal Noise}

In order to elucidate the potential benefits of multi-material coatings
 we will first describe briefly the model of thermal noise used in our calculations.
For Brownian thermal noise we start with \cite{Hong2013},
 and \cite{Evans2008} is our starting point for thermo-optic noise,
 though similar treatments can be found in \cite{Gorodetsky2008,Gurkovsky2010,Kondratiev2011}.

Since Hong et.al.\cite{Hong2013} conclude that changes in the ratio of shear to bulk mechanical loss
 do not significantly change the optimal coating design,
 and that photoelastic effects are relatively unimportant,
 we can simplify their result significantly by assuming that shear and bulk mechanical losses
 are equal, $\phi_M = \phi_{\rm bulk} = \phi_{\rm shear}$, and that the photoelastic effects can be ignored.
 (While not important for optimization, the ratio of shear and bulk losses impacts
  the level of Brownian thermal noise at the $\pm30\%$ level \cite{Hong2013}.)
The resulting equation for Brownian thermal noise is
\beq{brownian}
S_z^{Br} = \frac{4 k_B T}{\pi r_G^2 \omega}  ~ \frac{1 - \prat_s - 2  \prat_s^2}{Y_s} 
  \sum_j b_j d_j \phi_{Mj} \\
\eeq
 where the unitless weighting factor $b_j$ for each layer is
\beq{brownian_weight}
b_j = \frac{1}{1 -  \prat_j} 
\[  \( 1 - n_j \partderiv{\phi_c}{\phi_j} \)^2 \frac{Y_s}{Y_j} + 
\frac{(1 - \prat_s - 2  \prat_s^2)^2}{(1 + \prat_j)^2 (1 -  2 \prat_j)} \frac{Y_j}{Y_s} \] . \nonumber
\eeq
Under the assumption that the substrate and coating elastic parameters are equal
 ($Y_j \rightarrow Y_s$ and $\prat_j \rightarrow \prat_s$),
 and ignoring field penetration into the coating ($\partderiv{\phi_c}{\phi_j} \rightarrow 0$),
 $b_j \rightarrow 2$ for all layers.

For thermo-optic noise we use
\beq{thermo-optic}
S_z^{TO} = \frac{4 k_B T^2}{\pi r_G^2 \sqrt{2 \kappa_s C_s \, \omega}}
 \[ \abar_c d - \bbar \lambda_0 -  \frac{\abar_s}{C_s} \sum_j d_j C_j \]^2
\eeq
where
\beqa{thermo-optic_alpha_beta}
\abar_s &=& 2 (1 + \prat_s)  \alpha_s \\
\abar_c &=& \sum_{j = 1}^N \frac{d_j}{d} 
 \frac{1 + \prat_s}{1 - \prat_j} \[ \frac{1 + \prat_j}{1 + \prat_s} + (1 - 2 \prat_s) \frac{Y_j}{Y_s} \] \alpha_j \\
\bbar &=& \sum_{j = 1}^N \frac{d_j}{\lambda_0}
  \( \beta_j + \frac{1 + \prat_j}{1 - \prat_j} \alpha_j n_j \) \dcdp ~.\elabel{betaTR}
\eeqa
Note that the expression for $\bbar$ is slightly different from that of \cite{Evans2008},
 thanks to the correction by K. Yamamoto in chapter 8.2.5 of \cite{Harry2012}.

In this paper we make a number of simplifying assumptions
 ($\phi_M = \phi_{\rm bulk} = \phi_{\rm shear}$, no photoelastic effect),
 and we ignore several correction factors (thick coating correction \cite{Evans2008},
 finite size test-mass corrections \cite{Braginsky2003, Somiya2009}).
All of this is to keep the formalism simple enough that the results can be easily understood
 and evaluated, but it should not be taken to mean that these corrections cannot be applied
 to multi-material coatings.
Indeed, their application is expected to be a straight-forward if somewhat messy process.
 
%%%%%%%%%%%%%%%%%%%%%%%%%%%%%%%%%%%%%%%%%%%%%%%%%%%%%
\subsection{Reflection Phase}
\sslabel{ReflCoat}

In this section we summarize the model for coating reflectivity
 presented in Appendix B of \cite{Evans2008},
 as this calculation forms the basis for computing the
 coating phase sensitivity to mechanical and thermal fluctuations
 (e.g., $\tdcdp$ in eqns \eref{brownian} and \eref{betaTR}).
In the next section, we extend this computation to include
 distributed absorption in coating materials,
 which is an essential ingredient in the primary result of this paper.

%%%%%%%%
\begin{figure}[ht!]
\centering
\includegraphics[width=0.45\textwidth]{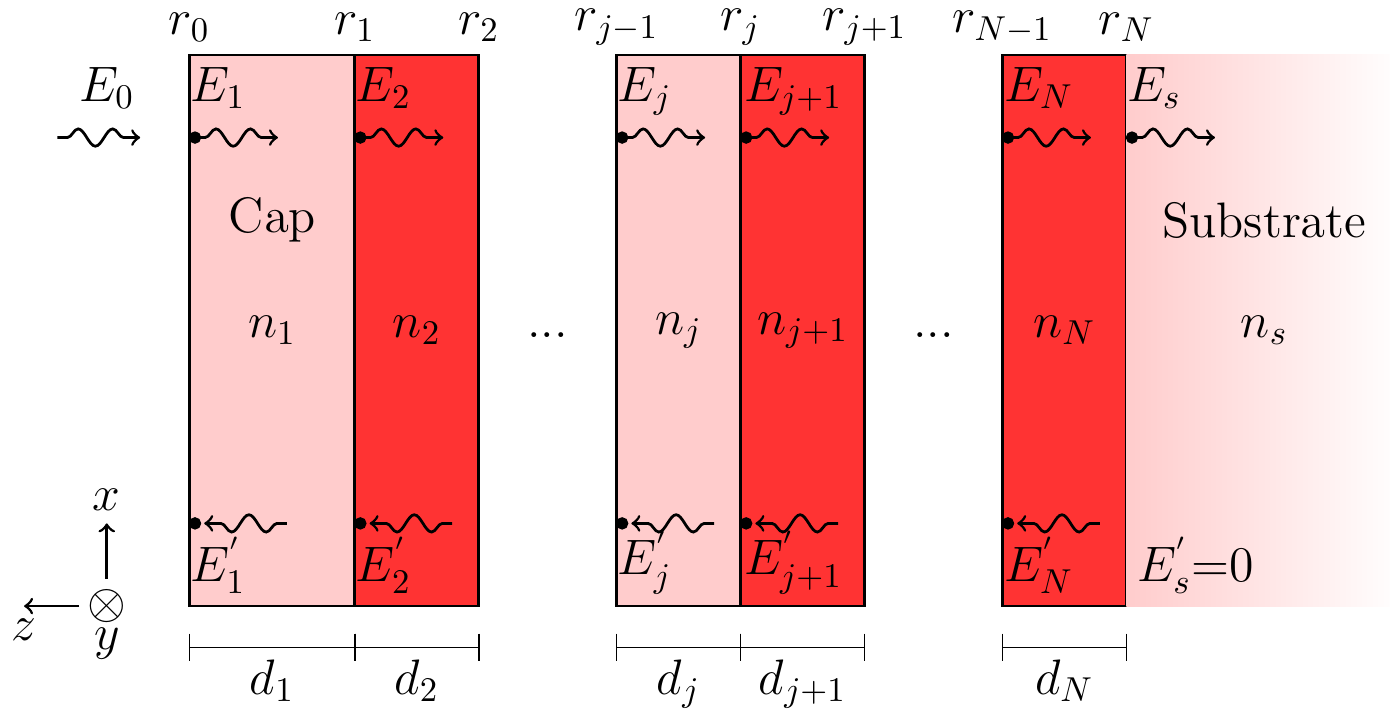}
\caption{The numbering of coating layers, interfaces and fields is shown above.
The z-coordinate is zero at the coating surface, and is positive moving
 away from the coating.
Note that the E-fields are evaluated just inside each coating layer
 (e.g., $E_1$ is evaluated at $z = -\epsilon$ in the limit of $\epsilon \rightarrow 0$). 
\flabel{CoatingLayers}}
\end{figure}
%%%%%%%%

As in \cite{Evans2008}, we express the reflectivity of the interface between two coating layers,
 seen by a field moving from layer $j$ to layer $j + 1$ (see figure \fref{CoatingLayers}), as
\begin{equation}
r_j = \frac{n_j - n_{j+1}}{n_j + n_{j+1}}.
\elabel{r_AB}
\end{equation}

%We number the layers in a coating in the order of increasing depth
% (i.e., the layer in contact with the vacuum is number $1$,
 % and the one in contact with the substrate is $N$)
By recursively combining the interface reflectivities $r_j$,
 we can find the reflectivity of layer $j$ and all of the layers between it and the substrate
\begin{equation}
\rbar_j = \frac{E_j}{E_j^\prime} = e^{i \phi_j} \frac{r_j + \rbar_{j + 1}}
 {1 + r_j \rbar_{j + 1}}.
\elabel{rbar}
\end{equation}
Note that while $\rbar_j$ includes the roundtrip propagation phase in layer $j$,
 it does not include the reflectivity of the interface between layer $j - 1$ and $j$.

The expression for $\rbar_j$ is recursive and the base case is the transition from
 the $N^{th}$ coating layer to the substrate,
\begin{equation}
\rbar_N = e^{i \phi_N} r_{N}
\end{equation}
 which can be evaluated with (\eref{r_AB}) using $n_{N+1} = n_s$.
Total coating reflectivity $\rbar_0$ is evaluated with the external vacuum
 acting as layer zero such that $n_0 = 1$.

The sensitivity of the coating reflection phase to a change in layer $j$
\begin{equation}
\dcdp
 = \imag{\frac{1}{\rbar_0} \partderiv{\rbar_0}{\phi_j}}
 = \imag{\partderiv{~ \log \rbar_0}{\phi_j}}.
\elabel{phi_refl}
\end{equation}
 is given by the recursion relation
\begin{equation}
\partderiv{\rbar_k}{\phi_j} =
\begin{cases}
 e^{i \phi_k} \frac{1 - r_k^2}{(1 + r_k \rbar_{k + 1})^2}
  \partderiv{\rbar_{k + 1}}{\phi_j} & k < j \\
 i \rbar_k & k = j \\
 0 & k > j
\end{cases}
\elabel{dr_k}
\end{equation}
 with the recursion starting at $k = 0$, progressing through increasing values of $k$,
  and terminating at $k = j$.

%%%%%%%%%%%%%%%%%%%%%%%%%%%%%%%%%%%%
\subsection{Optical Absorption}

Maintaining extremely low optical absorption in high-reflection coatings
 severely limits the choice of coating materials \cite{Beauville2004}.
The key idea behind this paper and its experimental counterpart \cite{Steinlechner14a},
 is that this stringent requirement need not be applied to all layers in the coating,
 but only to those near the surface which are dominantly responsible for the
 absorption of the coating.

To compute the depth dependence of optical absorption in a coating,
 we start by evaluating the electric field present in each layer of the coating
\beq{Efield_A}
E_{j + 1} = \frac{\sqrt{1 - r_j^2}}{1 + r_j \rbar_{j + 1}} E_j
\eeq
 (see figure \fref{CoatingLayers}).
This expression can be used iteratively to compute the field entering each
 coating layer given that the field entering the coating from the vacuum is
 $E_0 = \sqrt{2 P_0 / \pi \rG^2 c \epsilon_0}$, where $P_0$ is the power of the incident laser beam,
  while $c$ and $\epsilon_0$ are the speed of light and the permittivity of free space 
 (see \cite{Hong2013} or Appendix A of \cite{Kwee2014}).

The field at any point in a given layer will be the sum
 of the two counter propagating fields 
\begin{gather}
\elabel{Efield_2}
E(z_j, t) = \real{E_j e^{i (\omega t + k_0 n_j z_j)} + E_j^\prime e^{i (\omega t - k_0 n_j z_j)}} \\
\mbox{where} ~~ z_j = z - \sum_{k=1}^{j-1} d_k
\end{gather}
 such that $z_j = 0$ at the top of layer $j$, and $z_j = -d_j$ at the bottom.
 Optical absorption per unit length in a layer is assumed to be proportional to
 the time averaged field amplitude squared integrated over that layer,
 normalized by the power entering the layer and the layer thickness
\beqa{Efield_3}
\rho_j &=& \frac{2}{\abs{E_j}^2 d_j} \left< \int_{-d_j}^{0} E(z_j, t)^2 dz_j \right>_t\\
 &=& (1 + \abs{\rbar_j}^2 ) + 2 \frac{\fsin{k_0 n_j d_j}}{k_0 n_j d_j} \real{\rbar_j e^{i k_0 n_j d_j}} ~.
\eeqa
Note that the second term is zero for quarter-wave layers (i.e., with $k_0 n_j d_j = \pi / 2$),
 and that $\rho_j$ is constructed such that $\rho_j = 1$ for $\rbar_j = 0$
 (i.e., for a field propagating in the absence of a counter propagating field).

Using equation (\eref{Efield_A}) we can further relate the absorption in each layer to
 the total absorption coefficient for the coating  $a_c$, by
\beq{aCoat}
a_c = \sum_{j = 1}^{N} \rhobar_j a_j d_j ~~\mbox{where}~~
\rhobar_j = \frac{\abs{E_j}^2}{\abs{E_0}^2} \rho_j
\eeq
 and $a_j$ is the absorption per unit length of the material used in layer $j$.

Absorption loss in coatings is usually quoted as a single value, the total $a_c$,
 rather than an absorption per unit length for the coating constituents
 \cite{Abernathy2008,Flaminio2010}.
Using equation \eref{aCoat} we can convert absorption values in the literature
 into absorption per unit length.
Assuming that odd layers are \sil\ with negligible absorption,
\beq{aMaterial}
a_{X} = \frac{1}{a_c} \sum_{j = even}^{N} \rhobar_j d_j 
\eeq
which we use to compute the value for \tant\ presented in table \tref{materials}.

%%%%%%%%%%%%%
\begin{table}[h!]
\begin{center}
\begin{tabular}{ccc}
 symbol ~ & name  & unit or value\\
 \hline
 $\kB$ & Boltzmann's constant & \SI{1.38e-23}{J/K}\\
 $T$ & mean temperature & \SI{290}{K}\\
 $\lambda_0$ & vacuum laser wavelength & \SI{1064}{nm}\\
 $k_0$ & laser wavenumber $= 2 \pi / \lambda_0$ & \SI{5.9e6}{/m}\\
 $r_G$ & Gaussian beam radius ($1/e^2$ power) & \SI{6}{cm}\\
\hline
 $\omega$ & angular frequency & \si{rad/s}\\
 $S_z^{Br}$ & Brownian noise & \si{m^2/Hz} \\
 $S_z^{TO}$ & Thermo-optic noise & \si{m^2/Hz} \\
\hline
 $n$ & refractive index &  \\
 $\alpha$ & thermal expansion & \SI{1}{/K}\\
 $\beta$ & $\partderivT{n}{T}$ & \SI{1}{/K}\\
 $\kappa$ & thermal conductivity & \si{W/K m}\\
 $C$ & heat capacity per \textbf{volume} & \si{J/K m^3}\\
 $Y$ & Young's Modulus & \si{N/m^2}\\
 $\prat$ & Poisson ratio &  \\
 $\phi_M$ & mechanical loss angle &  \\
 $a$ & optical absorption &  \si{1/m}\\
\hline
 $d$ & coating thickness & \si{m} \\
 $z$ & depth in the coating (negative) &  \si{m} \\
 $P$ & laser power arriving at each layer & \si{W} \\
 $E$ & complex electric field amplitude & \si{N/C} \\
 $r$ & complex amplitude reflectivity &  \\
 $\rho$ & power absorption ratio &  \\
 $b$ & Brownian weight coefficient & 
\end{tabular}
\end{center}
\caption{Frequently used symbols for physical constants, the environment,
 material parameters, etc., are given above.
Material parameters that appear with a subscript refer to either the
 substrate material, subscript $s$, the coating, subscript $c$,
 or are indexed to a particular coating layer, typically with the variable $j$.
Note that $k$ is occasionally used as a local index for summation or
 recursion when $j$ is already in use.\tlabel{SymbolDefs}}
\end{table}
%%%%%%%%%%%%%

%%%%%%%%%%%%%
\vspace{10pt}
\begin{table}[ht]
\begin{center}
\begin{tabular}{cccccc}
 property ~ & ~ \sil ~ & ~ \tant ~& ~ $\rm MO_A$ ~& ~ $\rm MO_B$ ~&~ unit \\
 \hline
 $n$		& 1.45 	& 2.1		& 2.1 	& 3.0 	& 1 \\
 $\alpha$	& 0.51 	& 3.6		& 3.0		& 3.0		& \SI{e-6}{/ K}\\
 $\beta$	& 8 		& 14		& 10		& 10		& \SI{e-6}{/ K}\\
 $\kappa$	& 1.38 	& 33 		& 30		& 30		& \si{W / m K}\\
 $C$		& 1.64 	& 2.1 	& 2.0	 	& 2.0	 	& \si{MJ / K m^3}\\
 $Y$		& 72 		& 140	& 100 	& 70 		& \si{GPa}\\
 $\prat$	& 0.17 	& 0.23	& 0.2		& 0.2		&  1 \\
 $\phi_M$	& 0.4		& 3.8		& 1.0 	& 1.0 	& \num{e-4} \\
 $a$		& \num{e-3} 	& 2	& 10 		& 100 		& \si{ppm / \um} 
\end{tabular}
\end{center}
\caption{The values of material parameters used for all figures and examples.
The values for \sil\ and \tant\ are taken from \cite{Evans2008},
 while those of the hypothetical metal-oxides $\rm MO_A$ and $\rm MO_B$
 have been invented by the authors for use in examples
 presented in the text.\tlabel{materials}}
\end{table}
%%%%%%%%%%%%%

%%%%%%%%%%%%%%%%%%%%%%%%%%%%%%%%%%%%
\section{Example Coatings}
\sslabel{examples}

Given the coating model described in the previous section,
 and a pallet of possible coating materials,
 we can evaluate the impact of using more than two materials to
 make a coating.
In this paper we allow ourselves two hypothetical coating materials,
 metal-oxide A and B ($\rm MO_A$ and $\rm MO_B$) as a means of
 demonstrating the types of optimizations which can occur (see table \tref{materials}).

The coating examples presented in this section are designed to show
 how multi-material coatings can \emph{in principal} be used to produce
 low-noise coatings.
For a detailed application of this approach to three-material coatings
 involving amorphous silicon see \cite{Steinlechner14a}.
 
As a baseline, we start by computing the thermal noise seen by \SI{1064}{nm} light
 for a 20-layer coating, made of 10 \sil-\tant\ layer pairs or ``doublets''. 
The top layer, known as the ``cap'' has an optical thickness
 equal to half of the laser wavelength,
 such that $d_1 n_1 / \lambda_0 = \half$.
All of the deeper coating layers are quarter-wave with $d_j n_j / \lambda_0 = \tfrac{1}{4}$.
The results of calculations for this coating are shown in figures
 \fref{rho}, \fref{dcdp} and \fref{ctn}.
This coating transmits 0.1\% of the incident light power,
 and absorbs \SI{0.5}{ppm}.

%%%%%%%%
\begin{figure}[ht!]
\centering
\includegraphics[width=0.42\textwidth]{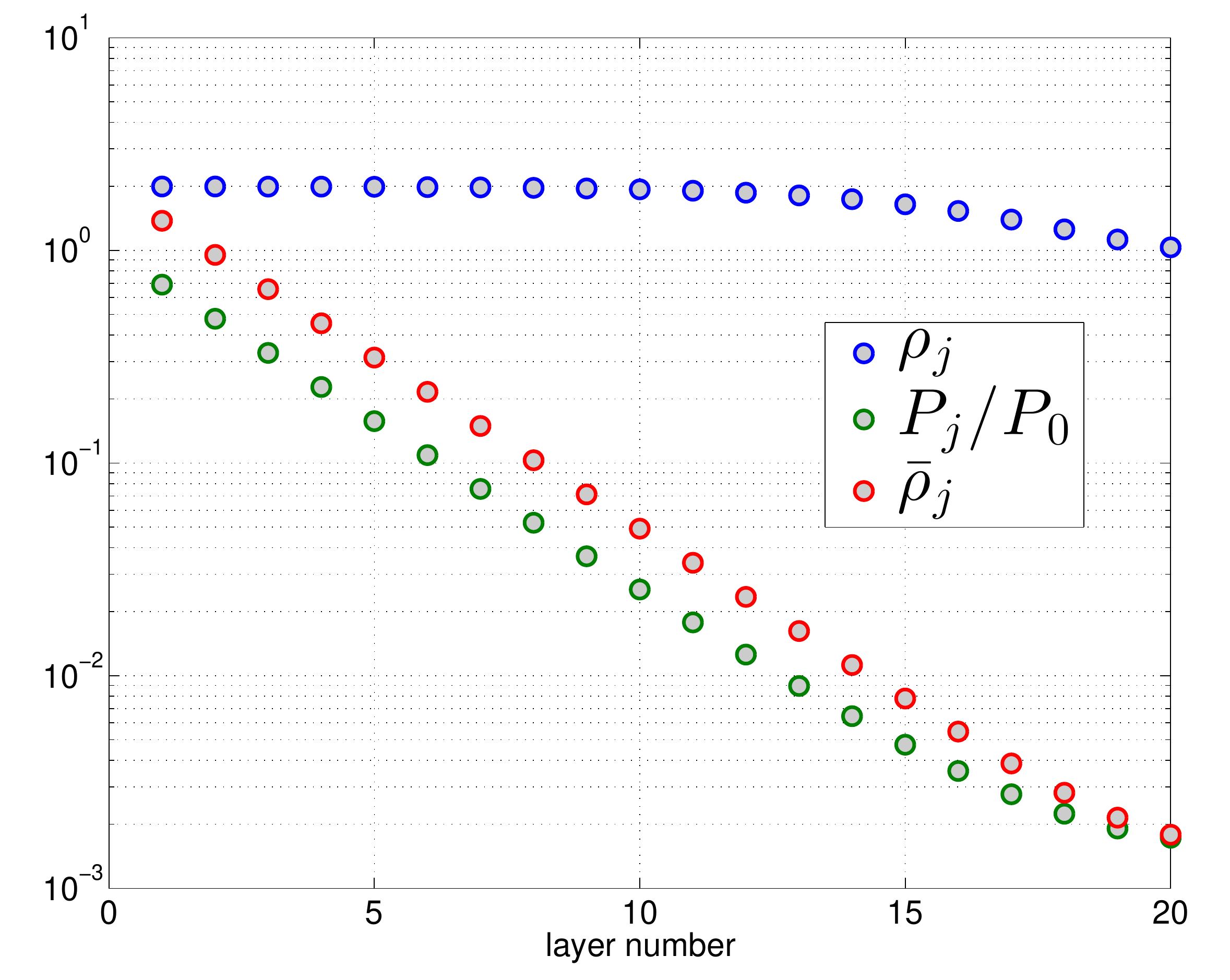}
\caption{The contribution of coating layers to the total coating absorption
 varies with depth in the coating.
Deeper layers make a smaller contribution,
 as seen by the rapidly decreasing value of $\rhobar_j$ with increasing $j$.
The two ingredients of $\rhobar$, absorption of each layer relative to the
 power arriving at that layer, $\rho_j$,
 and power attenuation as a function of depth $P_j / P_0$, are also shown
 (see eqns \eref{Efield_3}-\eref{aCoat}).
The coating used to generate this figure consists of 10 \sil\ - \tant\ quarter-wave doublets.
\flabel{rho}}
\vspace{-10pt}
\end{figure}
%%%%%%%%

For comparison, we can change the high-index material used below the top 3
 coating doublets to $\rm MO_A$, which is similar to \tant\ but is somewhat softer,
 has lower mechanical loss, and much higher absorption.
The lower Young's modulus makes a better match to the \sil\ substrate,
 and combines with the lower $\phi_M$ to reduce the Brownian noise
 of this coating to 70\% of the baseline coating.
The transmission of this coating is the same as the baseline,
 and the absorption is only slightly higher at \SI{0.6}{ppm}.

A more extreme example is a coating made of 4 \sil-\tant\ doublets,
 and 3 \sil-$\rm MO_B$ doublets.
This coating has less than 70\% of the Brownian noise of the baseline
 coating, and only \SI{0.8}{ppm} absorption.
The high refractive index of this material means that fewer and thinner
 layers are needed relative to \tant\ to produce the same transmission.
This, in combination with the good mechanical properties of this coating,
 more than compensate for its high absorption of \SI{100}{ppm/\um}.

%%%%%%%%
\begin{figure}[ht!]
\centering
\includegraphics[width=0.42\textwidth]{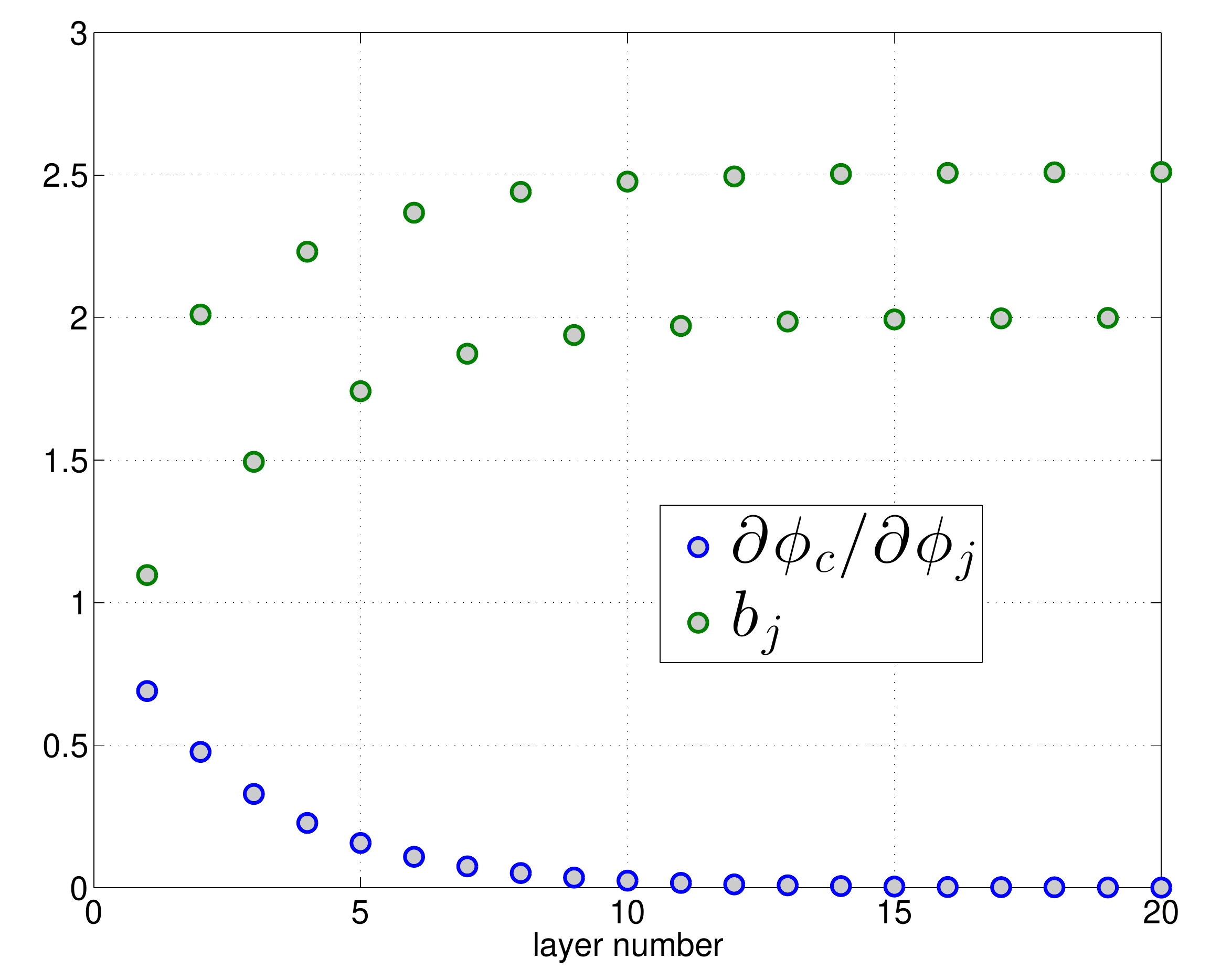}
\caption{The contribution of coating layers to Brownian noise increases with depth in the coating.
This figure shows the coating phase sensitivity to fluctuations in each layer, $\tdcdp$,
 and the weight of each layer in the total Brownian noise,
 $b_j$ (see equations \eref{brownian} and \eref{phi_refl}).
%Depending on the elastic properties of the coating materials, this may represent a significant
% reduction in the Brownian noise resulting from the first few layers of the coating.
%The coating used to generate this figure consists of 10 \sil\ - \tant\ quarter-wave doublets.
\flabel{dcdp}}
\end{figure}
%%%%%%%%

%%%%%%%%
\begin{figure}[ht!]
\centering
\includegraphics[width=0.42\textwidth]{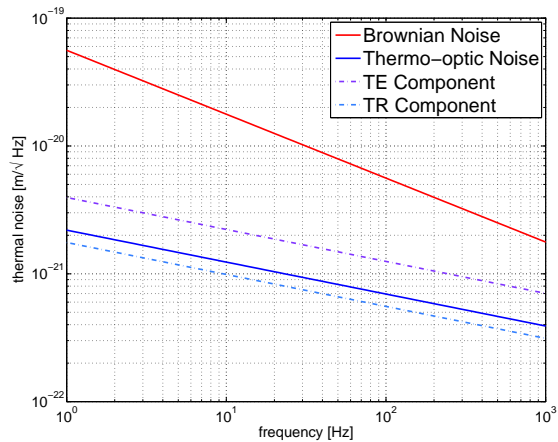}
\caption{Brownian thermal noise dominates in this example coating,
 in part thanks to the cancellation of the thermo-elastic (TE) and thermo-refractive (TR)
 components of thermo-optic noise.
The noise level of \SI{4.6e-21}{m/\rtHz} at \SI{100}{Hz} is a useful benchmark
 for the impact of coating thermal noise on gravitational wave detectors.
The coating used to generate this figure consists of 10 \sil\ - \tant\ quarter-wave doublets.
\flabel{ctn}}
\end{figure}
%%%%%%%%

%%%%%%%%%%%%%%%%%%%%%%%%%%%%%%%%%%%%
\section{Conclusions}

Precision optical measurements are increasingly limited by coating thermal noise,
 and much time and effort has been and continues to be spent in the search for better
 coating materials \cite{Kessler2012}.
In this work we suggest that the search for coating materials should not focus on finding
 a single material which satisfies all requirements, but rather a pallet of materials
 which together can be used to make coatings which satisfy all requirements.

While a single high-index, low absorption and low mechanical loss material would be ideal,
 the examples in this work show that a high-index material with low mechanical loss,
 but not necessary low optical absorption, will suffice to make lower noise coatings possible.
Since the material properties of a given coating layer depend not only on its constituents (e.g., doping),
 but also on the manufacturing process (e.g., annealing) a wide range of material properties
 have already been measured or are potentially accessible.

% If you have acknowledgments, this puts in the proper section head.
\begin{acknowledgments}
The authors gratefully acknowledge the support of the National Science
Foundation and the LIGO Laboratory, operating under cooperative
Agreement No. PHY-0757058. This paper has been assigned LIGO Document
No. LIGO-P1400206.
\end{acknowledgments}

%%%%%%%%%%%%%%%%%%%%%%%%%%%%%%%%%%%%%%%%%%%%%%%%%%%%%
%%%%%%%%%%%%%%%%%%%%%%%%%%%%%%%%%%%%%%%%%%%%%%%%%%%%%
\appendix

%%%%%%%%%%%%%%%%%%%%%%%%%%%%%%%%%%%%%%%%%%%%%%%%%%%%%
%\section{Coating Average Properties}
%\sslabel{CoatAvg}
%
%Optical coatings are made from alternating layers of materials with
% different refractive indices.
%For properties other than the refractive index,
% as long as the length scales involved ($\rdel$ and $\rG$) are large
% compared to the layer thickness (typically $< \lambda / 2$),
% we can use suitably averaged material properties to
% represent the coating.
%The equations given in this section are all taken from \cite{Evans2008},
% and are repeated here only for completeness and clarity.
%
%The total coating thickness
%\begin{equation}
%\elabel{d_sum}
%d = \sum_{k = 1}^{N} d_k.
%\end{equation}
%
%The heat capacity is a simple volume average,
%\begin{equation}
%\elabel{C_avg}
%C_\coat = \sum_{k = 1}^{N} C_k \frac{d_k}{d}
%\end{equation}
% while the average thermal conductivity involves the inverse
%\begin{equation}
%\elabel{K_avg}
%\kappa_\coat = \left( \sum_{k = 1}^{N} \frac{1}{\kappa_k} \frac{d_k}{d} \right)^{-1} .
%\end{equation}
%
%%%%%%%%%%%%%%%%%%%%%%%%%%%%%%%%%%%%%%%%%%%%%%%%%%%%%
\section{Relation to Other Works}
\sslabel{Relate}

The expressions in this work are related to those of \cite{Hong2013}  by
\beqa{}
\rbar_0 &=& \rho_{\rm tot} \\
2 \dcdp &=& \imag{\partderiv{\log \rho}{\phi_j}} ~\forall~ 1 \leq j \leq N \\
 &=& \imag{\epsilon_j / n_j} ~ \mbox{for their} ~ \beta_j = 0
\eeqa
 with our expression on the left of each equality and theirs on the right.
The factor of 2 results form our definition of $\phi_j$ as a \textbf{round-trip} phase
 in each layer, while theirs is a one-way phase.
 
We did, however, follow the convention of \cite{Hong2013} for the direction of the z-axis:
 normal to the surface of the coating and pointing \emph{into} the vacuum.
This represents a sign reversal relative to \cite{Evans2008},
 such that $\tdcdp$ is generally positive in this work as in \cite{Hong2013}.

An earlier treatment of Brownian thermal noise which included field
 penetration effects was performed in \cite{Gurkovsky2010}.
They based their computation on the coating reflection phase sensitivity to
 interface displacements (rather than layer thickness changes)
 and their notation is connected to ours by
\beq{}
n_j \dcdp - n_{j + 1} \partderiv{\phic}{\phi_{j + 1}} = \epsilon_j
\eeq
 again with our expression on the left of the equality and theirs on the right.

%%%%%%%%%%%%%%%%%%%%%%%%%%%%%%%%%%%%%%%%%%%%%%%%%%%%%
%%%%%%%%%%%%%%%%%%%%%%%%%%%%%%%%%%%%%%%%%%%%%%%%%%%%%

% Create the reference section using BibTeX:

\bibliography{papers}

\end{document}